\documentclass[aps,prl,twocolumn,superscriptaddress]{revtex4}

\usepackage[dvips]{graphics}

\begin{document}
\title{Evidence of spontaneous spin polarization in the two-dimensional electron gas}

\author{A.~R.~Go\~ni\footnote[1]{ICREA Research Professor, Barcelona, Spain}}
\affiliation{Institut f\"{u}r Festk\"{o}rperphysik, Technische
Universit\"at Berlin, Hardenbergstr. 36, 10623 Berlin, Germany}

\author{P.~Giudici}
\affiliation{Institut f\"{u}r Festk\"{o}rperphysik, Technische
Universit\"at Berlin, Hardenbergstr. 36, 10623 Berlin, Germany}

\author{F.~A.~Reboredo}
\affiliation{Lawrence Livermore National Laboratory, Livermore, CA 94551, USA}
\affiliation{Comisi\'on Nacional de Energ\'{\i}a At\'omica, Centro At\'omico
Bariloche, 8400 S.C.~de Bariloche, Argentina}

\author{C.~R.~Proetto}
\affiliation{Comisi\'on Nacional de Energ\'{\i}a At\'omica, Centro At\'omico
Bariloche, 8400 S.C.~de Bariloche, Argentina}

\author{C.~Thomsen}
\affiliation{Institut f\"{u}r Festk\"{o}rperphysik, Technische
Universit\"at Berlin, Hardenbergstr. 36, 10623 Berlin, Germany}

\author{K.~Eberl}
\affiliation{MPI f\"ur Festk\"orperforschung, Heisenbergstr. 1,
70569 Stuttgart, Germany}

\author{M.~Hauser}
\affiliation{MPI f\"ur Festk\"orperforschung, Heisenbergstr. 1,
70569 Stuttgart, Germany}

\begin{abstract}
Density-functional calculations using an exact exchange potential
for a two-dimensional electron gas (2DEG) formed in a GaAs single
quantum well predict the existence of a spin-polarized phase, when
an excited subband becomes slightly populated. Direct experimental
evidence is obtained from low temperature and low excitation-power
photoluminescence (PL) spectra which display the sudden appearance
of a sharp emission peak below the energy of the optical
transition from the first excited electron subband upon its
occupation. The behavior of this PL feature in magnetic fields
applied in-plane as well as perpendicular to the 2DEG indicate the
formation of spin-polarized domains in the excited subband with
in-plane magnetization. For it speaks also the strong enhancement
of exchange-vertex corrections observed in inelastic light
scattering spectra by spin-density excitations of a slightly
occupied first-excited subband.
\end{abstract}

\maketitle

\sloppy




The existence of a spin-polarized phase of the homogeneous
electron gas at densities between 10$^{20}$ and 10$^{18}$
cm$^{-3}$ and at zero magnetic field, although predicted years ago
\cite{bloch29a} and later well established from quantum
Monte-Carlo calculations \cite{ceper80a}, remains experimentally
unconfirmed. It has been suggested that a partially polarized
fluid state \cite{ceper99a} might be at the origin of the
high-temperature weak ferromagnetism observed recently in La-doped
calcium hexaboride \cite{young99a}. In this case, however, the
itinerant magnetism is still a matter of debate, for there might
be an alternative explanation in terms of point defects
\cite{monni01a}. Modulation-doped semiconductor heterostructures
offer a unique opportunity to study the many-body behavior of
dilute electron systems due to the extremely long mean-free paths
that can be achieved with current growth techniques. Experiments
carried out so far, using high-mobility electron gases formed in
GaAs/AlGaAs samples, exhibit, upon depletion, the metal-insulator
transition at about $2\times 10^{10}$ cm$^{-2}$
\cite{kravc94a,ernst94a,ilani01a} due to the poor screening of the
disorder potential of the ionized remote donors. In two dimensions
and without many-body screening, this transition should take place
for any degree of disorder \cite{abrah79a}. Localization effects
thus hamper further search for low-density phases of the
homogeneous two-dimensional electron gas (2DEG). Here we followed
a different approach by raising the Fermi level in a single
quantum well (SQW) structure in order to obtain a slightly
populated excited subband. By means of photoluminescence (PL)
spectroscopy and inelastic light scattering we were able to assess
the properties of the dilute component corresponding to the second
subband but having the 2DEG always in a metallic state because of
the charge present in the ground-state subband.

This Letter reports on possible ferromagnetic order in a
modulation-doped GaAs SQW, as predicted by self-consistent
density-functional calculations using an exact exchange potential
for the 2D electron system \cite{rebor03a} and as determined from
PL and Raman measurements at low temperatures and excitation
powers. Exchange terms of the Coulomb interaction lead to a
thermodynamical instability of the 2DEG formed in the quantum
well, as soon as the Fermi energy becomes resonant with the bottom
of the first-excited subband \cite{gonix02a}. At this first-order
phase transition the electron gas component corresponding to the
excited level breaks into spin-polarized domains with different
in-plane magnetization, as expected for planar ferromagnets to
minimise the stray field. Evidence for parallel spin alignment is
obtained from PL measurements of the light polarization at low
magnetic fields and from inelastic light scattering by
spin-density excitations involving electronic transitions between
the first and second excited subbands.


A 2DEG with a mobility of $8\times 10^5$ cm$^2$/Vs and a density
of $6\times 10^{11}$ cm$^{-2}$ at 4.2 K forms in a
modulation-doped GaAs/Al$_{0.33}$Ga$_{0.67}$As quantum well grown
by molecular-beam epitaxy. We present results of two samples with
25 and 30 nm well thickness. The 2DEG is contacted from the
surface by In alloying in order to apply a dc bias between it and
a metallic back contact. For the 25-nm-wide well and without bias
only the lowest subband is occupied with Fermi energy
$E_{F}\approx $~25 meV. The energy separation to the second
subband is $E_{01}\approx $~28 meV. For the wider well sample
there is a small occupation of the first-excited level at zero
voltage. Photoluminescence spectra were excited using low power
densities in the range of 0.5 to 5 W/cm$^2$ and recorded with
optical multichannel detection. Magneto-PL measurements were
carried out in Faraday as well as Voigt geometry by placing the
sample into the cold bore of a 5 T split-coil magnet. Inelastic
light scattering spectra were recorded in backscattering geometry
using a line focus in order to avoid heating of the 2DEG at the
laser powers required.


Band-structure calculations of the dilute 2DEG phases were performed in the
framework of density-functional theory (DFT) \cite{kohnx99a}. Implementation of
DFT requires non-trivial approximations in the exchange and correlation
contributions to the total energy. In the widely used local-density
approximation (LDA), both exchange and correlation potentials correspond to
results obtained for the homogeneous 3D electron gas. A key ingredient in our
calculations is the exact treatment of the exchange energy \cite{rebor03a},
whereas the remaining (small) correlation contribution to the total energy is
calculated within LDA. We also approximately accounted for low-temperature
effects in the exchange potential. Results obtained by solving the
self-consistent Kohn-Sham equations corresponding to the experimental sample
geometry are summarized in Fig. 1. For a given electric field which mimics the
applied bias, we calculate the total density $n$ and a set of energy eigenvalues
$E^{\sigma}_{i,\alpha}$ and eigenfunctions $\xi^{\sigma}_{i,\alpha}$, where
$i$=0,1 is the subband index, $\alpha$=PI, PII, F the phase index and $\sigma
=\uparrow,\downarrow$ the spin component. Depending on total density there are
three possible configurations: a) The paramagnetic phase denoted PI, where only
the ground subband is occupied with equal population for spin-up and down
components. This configuration is stable until $E_F^{PI} = E_{01,PI} =
E_{1,PI}-E_{0,PI}$. b) Phase PII, with both the first and second subbands
paramagnetically populated. This configuration is a self-consistent solution at
high densities, where $E_F^{PII}=E_{01,PII}/2 + \pi\hbar^2n/2m^*$, where $m^*$
is the electron effective mass. c) A ferromagnetic phase F for which the lowest
subband remains paramagnetic but only one spin component of the second subband
(say $\uparrow$) is occupied and the other empty.
This solution exists for $E_F^F\geq E^\uparrow_{01,F}$ but collapses towards the
PII configuration when $E_F^F=E^\downarrow_{01,F}$ because of the kinetic energy
cost to maintain a single spin population in the second subband.

\begin{figure}
\begin{center}
\resizebox{0.70\columnwidth}{!}{
\includegraphics*{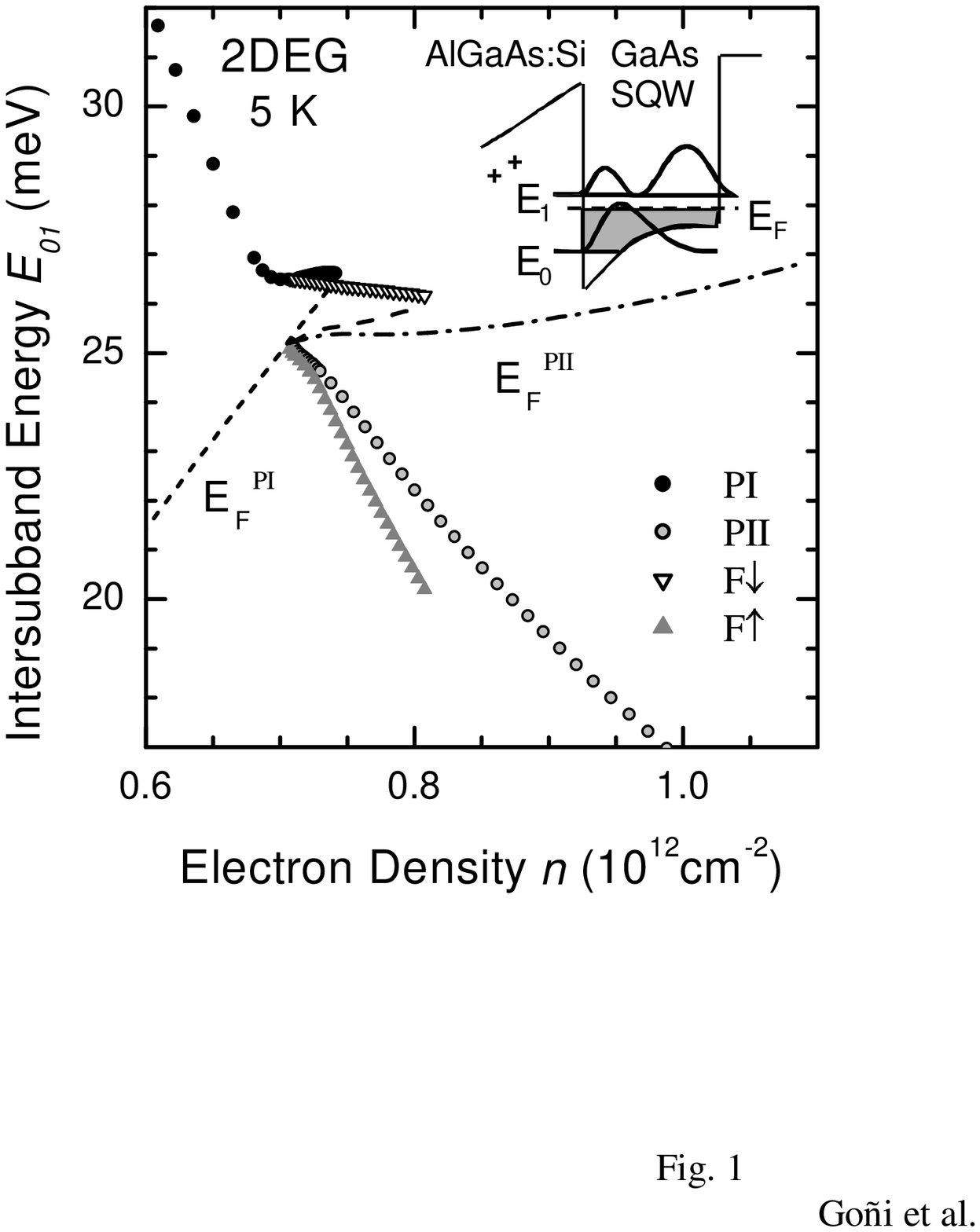}
}
 \end{center}
\caption{Dependence on total density $n$ of the intersubband
energy $E_{01}$ calculated within exact-exchange theory for a
25-nm-wide SQW at 5 K. Different symbols correspond to the
paramagnetic solutions with one (PI) and two (PII) occupied
subbands and the ferromagnetic phase (F), as indicated. Lines
indicate the position of the Fermi level in each phase. The inset
shows a sketch of the conduction band profile of the SQW. }
\label{theo}
\end{figure}

Figure 1 shows two remarkable effects, both beyond the reach of LDA: The abrupt
decrease of the intersubband spacing $E_{01}$ at about $0.72\times10^{12}$
cm$^{-2}$ when transforming from phase PI to PII and the very existence of the
spin-polarized phase F. The sudden renormalization of $E_{01}$, which is a
consequence of {\it intersubband} exchange terms, has been discussed in Ref.
\cite{gonix02a}.    For an incipient population of the excited subband the leading
exchange contribution to the 2DEG compressibility is negative, such that the
electron system becomes thermodynamically unstable against the shift of a
macroscopic amount of charge from the ground to the first-excited subband. When
$E_F=E_{01}$ electrons start to be transferred from states at the Fermi surface
of the lowest subband to Brillouin-zone center states in the excited 2D level.
This leads to the discontinuities observed in PL associated with the spontaneous
breaking of translational invariance.

In contrast, spin polarization is favored by {\it intrasubband} exchange terms
at low densities of the excited subband $n_1 = n_{1\downarrow}+n_{1\uparrow}$,
yielding a negative non-linear contribution that goes as $-n_{1\sigma}^{3/2}$.
With $n_{1\uparrow,\downarrow}=\left(\frac{n_1}{2}\mp\delta n_1\right)$ it is
given by
\begin{equation}
n_{1\uparrow}^{3/2}+n_{1\downarrow}^{3/2} \approx
2\left(\frac{n_1}{2}\right)^{3/2}\cdot\left[ 1+\frac{3}{2}\left(\frac{\delta
n_1}{n_1}\right)^2\right].
\end{equation}
Thus, the system gains energy by unbalancing the spin population of the second
subband. The last term in Eq. (1) is an exchange-hole energy gain in the excited
subband, which reduces the overall Coulomb repulsion in the 2DEG due to the
Pauli principle. With increasing $n_1$, positive contributions from the kinetic
and Hartree energies overcome the intrasubband exchange part and the F phase
collapses towards the paramagnetic PII phase.

\begin{figure}
\begin{center}
\resizebox{0.8\columnwidth}{!}{
\includegraphics*{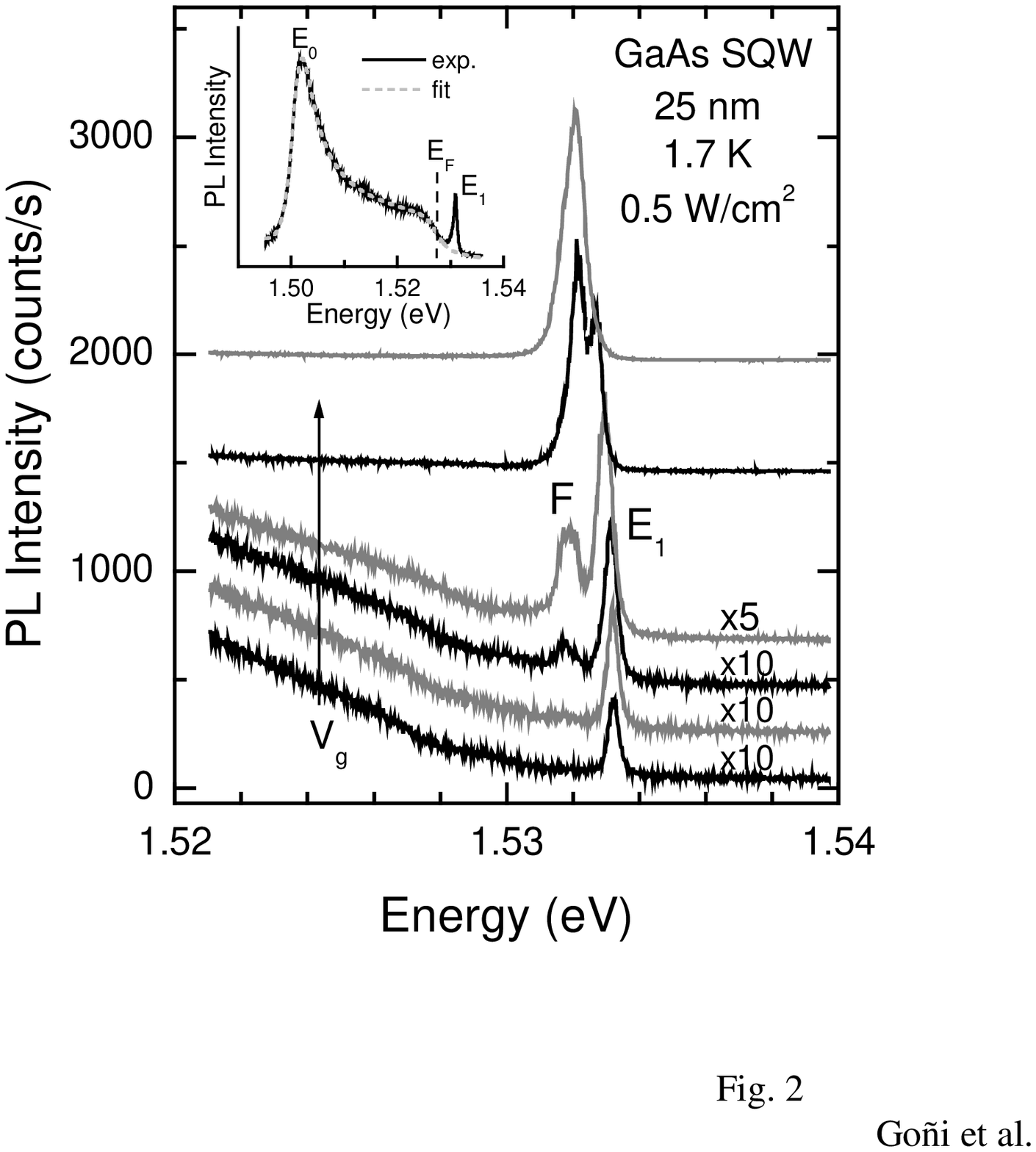}
}
 \end{center}
\caption{PL spectra of the 25-nm-wide SQW in the energy region of
the $E_1$ transition for different gate bias V$_g$ taken at 1.7 K
with 0.5 W/cm$^2$ power density. As an example, the inset displays
a full-range PL spectrum (solid curve) and the fitting function
(dashed line) used for lineshape analysis \cite{ernst94a}. }
\label{PLspec}
\end{figure}

Direct evidence of the existence of the F phase is obtained from
PL spectra recorded at low temperatures using very low laser
excitation. Figure 2 shows a set of spectra of the 25-nm-wide SQW
taken at 1.7 K and 3.5 $\mu$W laser power with different gate
voltage. As the Fermi level is raised towards resonance with the
second subband, an additional peak becomes apparent at about 1.2
meV below the energy $E_1$ of optical recombination processes from
the first excited electron subband. The F feature literally pops
up in PL, increasing in intensity while merging together with the
$E_1$ peak before its abrupt renormalization takes place. We
emphasize that peak F is only observable if the lattice
temperature is kept below 5 K and/or the power density does not
exceed 10 W/cm$^2$. On the contrary, at higher excitation levels
the $E_1$ peak instead of disappearing displays just a slight line
broadening and a strong increase in intensity. We also note that
close to the PI-PII phase transition the 2DEG is very unstable, a
situation which is reflected in the uncontrolled change of the
Fermi level, and thus of the PL spectrum, even without varying the
external voltage. In fact, the spectra of Fig. 2 correspond to
snapshots of 1 s during cycles where the electron system goes back
and forth having successively different occupations of the second
subband. The first order of the transition manifests itself in the
transformation of the $E_1$ lineshape from Lorentzian to Gaussian
and back to Lorentzian during the process of renormalization. Such
a change from homogeneous (lifetime) line broadening to an
inhomogeneous one (level distribution) is consistent with the
formation of density domains in the second subband during the
phase transition. The F feature is then associated with the
recombination of electrons within the domains, whereas the peak
$E_1$ arises from hot luminescence coming from regions of the SQW
plane where the second subband is still empty.

The energy position of the $E_1$ and F peaks measured at 2 K with 20 $\mu$W is
plotted in Fig. 3 as a function of the Fermi energy. When the Fermi energy
equals $E_1$, an abrupt subband renormalization by about 2 meV occurs.
Afterwards, one finds the 2DEG in the paramagnetic phase PII and the variation
of the subband energy with bias, i.e. Fermi level, is continuous but exhibits a
clear hysteresis. The stability range of the F phase comprises Fermi energies
only within 3 meV below $E_1$, and in this region the energy of the F peak is
almost independent of the Fermi level. This is a very important point, for it
rules out an explanation of our observations in terms of the Fermi-edge
singularity \cite{chenx92a}.

\begin{figure}
\begin{center}
\resizebox{0.80\columnwidth}{!}{
\includegraphics*{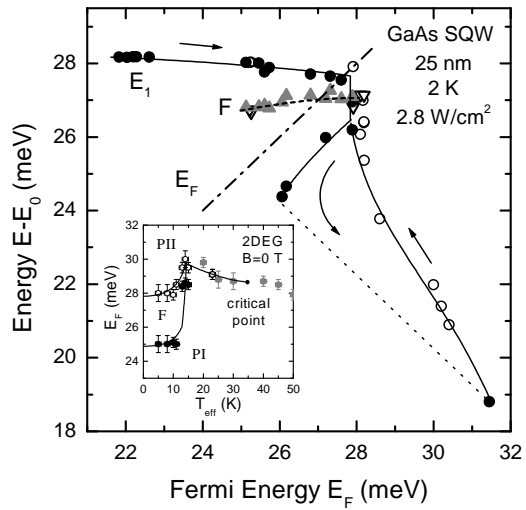}
}
 \end{center}
\caption{Energy position of the $E_1$ and F peaks referred to $E_0$, the bottom
of the ground subband, versus Fermi energy. The corresponding electron
temperature is 10 K. Solid (open) symbols correspond to data taken with
increasing (decreasing) bias. Solid lines are a guide to the eye. The inset
shows the phase diagram of the 2DEG at zero magnetic field obtained from PL
measurements. Grey data points are results of Ref. \cite{gonix02a} obtained at
high power densities.}
\label{energy}
\end{figure}

The inset to Fig. 3 illustrates the phase diagram of the 2DEG
conceived from the sets of data obtained at different temperatures
and laser powers. Effective electron temperatures were determined
from the width of the Fermi edge in PL spectra. Below a sample
temperature of 25 K, the temperature of the 2DEG is mainly given
by the excitation-power level. The lowest value of 5 K is obtained
with the sample immersed in superfluid helium and using only
around 1 $\mu$W of laser power. The region in the phase diagram
representing the stability range of the F phase corresponds to the
Fermi levels and temperatures for which the peak F is apparent in
PL. At 13 K, where the F phase collapses into the PII one, there
is a triple point. We note that this temperature corresponds to
the formation energy of the F phase, as given by the difference
between $E_1$ and F peak energy. As reported earlier
\cite{gonix02a}, there is a critical point at T$_C$=35 K. For
completeness we point out that the same phenomenology has been
observed in three other SQW samples with different well widths.

Having experimentally proven the existence of an additional phase
of the 2DEG, we now present the evidence for ferromagnetic order
stemming from magneto-PL measurements. We consider first the case
of magnetic fields applied perpendicular to the plane of the SQW
(Faraday geometry). Whereas the $E_1$ peak splits into two Zeeman
components with clearly different circular polarization
$\sigma^+$, $\sigma^-$ with the expected gyromagnetic factor
\cite{snell92a}, the F peak does not show any splitting at all nor
any degree of circular polarization. On the contrary, its
intensity decreases linearly with field, disappearing at about 1.5
T. Such behavior is compatible with the assumption that the
magnetization within the density domains is randomly, but
in-plane, oriented. A magnetic field in the perpendicular
direction destroys the domains since B$_\perp$ significantly
reduces the wave-function overlap in the second subband and thus
the intrasubband exchange interaction.

For magnetic fields applied along the 2DEG plane the intensity of
the F peak remains essentially constant. In contrast, its degree
of linear polarization defined as $\gamma = (I_V-I_H)/(I_V+I_H)$
increases linearly with B$_\parallel$ by 30\% in the range from
zero to 0.6 T. Here $I_V$ and $I_H$ are the intensities of the F
peak for vertical and horizontal linear polarization,
respectively, normalized to the intensity of the PL spectrum in
the energy region just underneath the Fermi edge, where the
emission is expected to be independent of its polarization. The
field is aligned along the horizontal direction, but rotated
90$^\circ$ with respect to the direction of light collection
(Voigt configuration). In this geometry, light emitted with
circular polarization along the field is detected as being
vertically polarized. Hence, the observed linear increase in
$\gamma$ can be ascribed to the successive orientation of the
domain magnetization parallel to the external field. Thus,
$\gamma$ is given by the Brillouin function $B_J$ which in the
low-field limit yields \cite{craik95a}
$\gamma = (J+1)g\mu_BB_\parallel/(3k_BT_{eff})$.
Here is $J$ the average total spin of the domains in units of
$\hbar/2$, $g$ =-0.44 is the gyromagnetic factor of GaAs
\cite{snell92a}, $\mu_B$ is the Bohr magneton, $k_B$ is the
Boltzmann constant, and $T_{eff}$ is the electron temperature.
From the measured slope for $\gamma$ we infer an average number of
$^1/_2$-spins, i.e. electrons, within a domain of about 50 (25 nm
SQW). For a typical electron density of $5\times10^{10}$ cm$^{-2}$
corresponding to the jump at the PI-PII phase transition the mean
radius of a domain should be around 200 nm.

\begin{figure}[h]
\begin{center}
\resizebox{0.65\columnwidth}{!}{
\includegraphics*{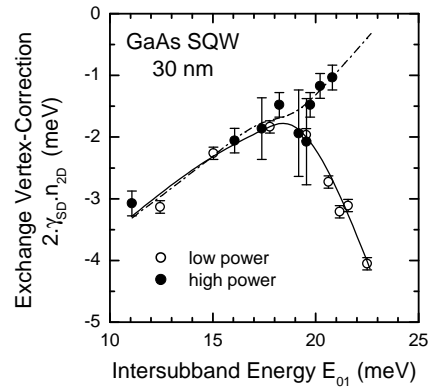}
}
 \end{center}
\caption{Exchange-vertex correction to the spin-density excitation
involving electronic transitions between subbands '1' and '2' as a
function of the intersubband spacing between the two lowest
subbands of a 30-nm-wide quantum well. Open(closed) symbols
correspond to measurements using low(high) power densities, i.e.
the effective electron temperature being 5 and 15 K, respectively.
Curves are a guide to the eye.} \label{raman}
\end{figure}

Another striking result which speaks for a spin polarization of
the dilute electron system in subband '1' is obtained from
inelastic light scattering. Collective spin-density (SD)
excitations of the 2DEG are known to be shifted down in energy
from the intersubband single-particle (SP) gap by many-body
effects, the so-called excitonic shift. It represents the vertex
correction due to exchange-correlation terms of the Coulomb
interaction \cite{pincz89a}. The determination of its magnitude is
straightforward from the energies of the excitations
$\hbar\omega_{SP}$ and $\hbar\omega_{SD}$ measured in depolarized
light scattering spectra, according to \cite{pincz89a}:
$-2\gamma_{SD}n_{2D} = \frac{\omega_{SD}^2 -
\omega_{SP}^2}{\omega_{SP}}$. Here $\gamma_{SD}$ represents the
exchange-correlation energy and $n_{2D}$ is the electron density
of the occupied subband. Figure \ref{raman} shows the values of
the exchange vertex correction to the spin-density excitation
associated with electronic transitions between the first and
second excited subbands of a 30-nm-wide SQW. In the graph, the
first-excited subband is being filled for decreasing subband
spacing $E_{01}$. Open symbols correspond to data obtained at a
very low power density such that the F peak is apparent in PL. For
comparison the full circles represent the high-power case, where
the effective electron temperature is too high for the
spin-polarized phase to exist (peak F is no longer observable).
The exchange vertex correction exhibits a strong enhancement at
low electron temperatures and very low filling of subband '1',i.e.
in the stability range of phase F, whereas at high power levels it
extrapolates to almost zero. The results are independent of laser
power, when the renormalization of $E_1$ sets in. As demonstrated
earlier for fractional quantum Hall states with odd filling factor
\cite{pincz92a} and in GaAs quantum wires in magnetic fields
\cite{gonix93a}, this is clear indication of spin polarization.
The cost in exchange energy to flip a spin in a ferromagnetically
ordered state exceeds many times that of the nonmagnetic case. We
emphasize that such an enhancement of vertex corrections is
clearly observed for the 1--2 electronic transitions rather than
the '0, 1' subbands, for in the latter case most of the electrons
involved in the spin excitation are in a paramagnetic state.

{\normalsize We thank F. Guinea for inspiring discussions and W.
Dietsche for assistance with the sample design and growth. F.A.R.
acknowledges the auspices of the U.S. Dept. of Energy at the
University of California/Lawrence Livermore National Lab under
contract no. W-7405-Eng-48. P.G. thanks the DAAD for financial
support. This work is supported in part by the DFG in the
framework of Sfb 296. }


\end{document}